\documentclass[a4paper]{paper} 
\usepackage{url,lineno}

\usepackage{geometry} 
\usepackage{graphicx} 
\usepackage{amsmath}
\usepackage{booktabs} 
\usepackage{array}    
\usepackage{verbatim} 

\begin{document}

\title{Phase Diagram of Spiking Neural Networks}
\author{Hamed Seyed-allaei  \\ {\it hamed@ipm.ir}} 
\institution{School of Cognitive Science, Institute for Research in Fundamental Sciences (IPM),     
\\Tehran, Iran.}

\maketitle

\begin{abstract}
In computer simulations of spiking neural networks, 
often it is assumed that every two neurons of the network are connected by a probability of 2\%, 
20\% of neurons are inhibitory and 
80\% are excitatory. 
These common values are based on experiments, observations, 
and trials and errors, 
but here, 
I take a different perspective, inspired by evolution, I systematically 
simulate many networks, each with a different set of parameters, 
and then I try to figure out what makes the common values desirable.
I stimulate networks with pulses and then measure their: 
dynamic range, dominant frequency of population activities, total duration of activities, 
maximum rate of population and the occurrence time of maximum rate.
The results are organized in phase diagram. 
This phase diagram gives an insight into the space of parameters -- excitatory to inhibitory ratio, 
sparseness of connections and synaptic weights. 
This phase diagram can be used to decide the parameters of a model.
The phase diagrams show that networks which are configured according to the common values, 
have a good dynamic range in response to an impulse and  
their dynamic range is robust in respect to synaptic weights, and 
for some synaptic weights they  oscillate in $\alpha$ or $\beta$ frequencies, 
even in absence of external stimuli.
\end{abstract}

\section{Introduction}


A simulation of neural network consists of model neurons that interact via network parameters; 
The model is a set of differential equations that describes the behaviors of neurons and synapses, 
captured by years of electro physiological studies \cite{Nicholls2012}. 
A model can be as simple as Integrate-and-fire or as realistic as Hodgkin-Huxley \cite{Sterratt2011a}. 
Realistic models generate reliable results, but they are computationally expensive.
On the other hand, simple models consume less computer resources, 
at the cost of loosing details and biological plausibility;
therefore, there is a delicate balance between computational feasibility and biological plausibility, 
Fortunately, 
there are many reviews and books that compare different models 
\cite{Sterratt2011a, Herz2006a, Izhikevich2004a} and neural simulators \cite{Brette2007b}.
They help us to find a suitable model of a single neuron, 
set its parameters, and get reliable results in a reasonable time.

No matter how much a neural model is detailed and efficient, 
A carelessly constructed networks of perfect neural models, is misleading. 
It only produces wrong biologically plausible results, in a reasonable time.
Unfortunately, network parameters are a lot more ambiguous.

There are four essential parameters in a neural network simulation 
that I investigated in this work - 
the ratio of excitatory neurons to inhibitory neurons, 
the topology and sparseness of connections, inhibitory and excitatory synaptic weights.
To make it more general and realistic, one may consider 
the values follow a distribution. 
This distribution function may change the behavior.

There are experimental works that reads network parameters from nature 
\cite{Lefort2009, Ko2011, Brown2009, Perin2011, Binzegger2004, Feldmeyer2012}. 
For example, {\it Lefort et al} studied the share of inhibitory neurons in each layer of 
C2 barrel column of mouse \cite{Lefort2009}. 
They reported that 100, 16, 10, 8, 17, 17 and 9 percent of layer 1, 2, 3, 4, 5A, 5B and 6 are inhibitory neurons, 
and in total, 11 percents of the column is made of inhibitory neurons. 
They have also studied the synaptic connections of excitatory neurons,
they reported neurons of each layers 2, 3, 4, 5A, 5B and 6 is connected 
with the probability of 9.3, 18.7, 24.3, 19.1, 7.2 and 2.8 percents to the neurons of the same layer.
Moreover, they have studied synaptic weights by measuring unitary Excitatory Post Synaptic Potential (uEPSP),  
the mean uEPSP for connections inside a layer are 0.64, 0.78, 0.95, 0.66, 0.71 and 0.53 mV for layer 2 to 6 
but the maximum values are much larger, 3.88, 2.76, 7.79, 5.24, 7.16 and 3 mV for the same layers respectively.

The actual values in computer simulations usually are less diverse, 
In many computational studies, 20\% of the neurons are inhibitory,  
the network topology is a random graph with sparseness of 2\% and
there are only a couple of fixed synaptic weights for inhibitory and excitatory neurons.
Despite their simplicity, they observe many interesting phenomena like oscillations, synchrony, modulations 
\cite{Song2000, Brette2007b, Goodman2008, Buice2009, Akam2010, Mejias2012, Augustin2013} and even psychological disorders \cite{Bakhtiari2012}.
There are also many works that show the results are not actually sensitive to the parameters 
\cite{Prinz2004, Marder2007, Marder2011, Marder2011a, Gutierrez2013}.

Why do this parameters appear in nature? 
What are their evolutionary advantages? 
How much is the result of a simulation robust in respect to the parameters?
How does one decide about the network parameters in a simulation?
Which other parameters might give the same results?
These are few question that come to mind, while dealing with properties of neural networks, 
and they could be easily answered by looking at a phase diagram of neural networks, 
that shows the behavior of a network in respect to its parameters.
Like the phase diagram of water, tells us about the properties of water in different pressures and temperatures.
But such a phase diagram does not exist for spiking neural networks, yet. 
First, because of the large amount of computation needed and second, 
because of the ambiguity of the concept of phase or state in neural networks.

To answer these question, I could use sophisticated evolutionary algorithms, 
like the study of evolution of intelligence by \cite{McNally2012a},
however, I used a simple brute force search in the space of parameters, 
so I can also generate a phase diagram, similar to the work of \cite{Roy2013}.

\section{Method}

I did a brute force search in the space of parameters, 
by simulating a population of 1000 neurons, let say a part of a single layer of cortex, 
that receives excitatory signals from previous layer of cortex or an external stimuli, 
and then sends excitatory signals to the next layer. 
There are lateral connections among all types of neurons (excitatory and inhibitory) in a single layer, 
but only excitatory neurons interconnects different layers. 
I stimulated this network and watched its response. 
This was repeated over a wide range of parameters. 
At the end I calculated fitness of each set of parameters and then investigated the 
overlap of the best parameters according to my simulation with the nature's choices
(or computational neuroscientists' choices). 

To do any sort of evolutionary analysis, first we need to agree on something to optimize --a fitness function,
it  can be the network capacity to learn and store information \cite{Barbour2007},
or  it can be the speed and the quality of transmitted signals \cite{Chklovskii2002}.
But here, to define the fitness function, I remind you of three obvious evolutionary facts:
\begin{enumerate}
\item High {\em Dynamic Range} is evolutionary favored, 
an animal that can see during days and nights, has a higher survival chance than an animal which can see only during days. 

\item Small {\em Just-Noticeable Difference} is evolutionary favored, 
an animal which can discriminate more grey levels or colors, has a better chance to find its preys and foods.

\item Low {\em Energy Consumption} is also favored, an energy efficient animal, 
consumes less food and therefore has a better chance to survive \cite{Niven2008}.
\end{enumerate}
\begin{figure}
\centering
\includegraphics[width=8cm]{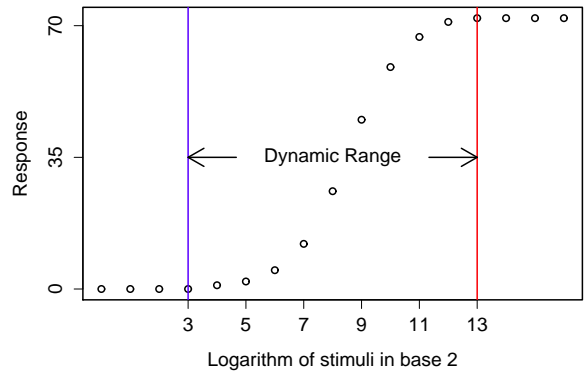}
\caption{This figure demonstrates a typical tuning curve and its dynamic range. 
Two consecutive stimuli are different by a factor of 2, also known as one {\it stop}.}
\label{fig:DynamicRange}
\end{figure}

Dynamic Range is the ratio of the largest possible value of stimuli to its smallest perceivable value. 
Just-Noticeable Difference is the smallest detectable difference  between two sensory stimuli,
which is often roughly proportional to the magnitude of the stimulus.
The ratio of Just-Noticeable Difference to the magnitude of the stimulus is known as Weber constant \cite{Dayan2001}.

Both Dynamic Range and Just-Noticeable are well presented in logarithmic scale, 
but instead of $\log_{10}$ and dB, from now on, 
I will use $\log_2$ and {\it stop} which are common in Optics and Photography (FIG. \ref{fig:DynamicRange}).

Here we are left with a neural network and three concepts to play with -- 
Dynamic Range, Just-Noticeable Different and Energy Consumption. 
To keep it simple, I only measure Dynamic Range while there are upper bounds for the other two. 
I require Just-Noticeable Different to be smaller than one {\it stop}, thus, 
the calculated Dynamic Range is a lower bound for the real Dynamic Range, 
and I require the Energy Consumption to be zero in absence of stimulus, 
which puts an upper bound for the Energy Consumption. 
Moreover, energy consumption can be a determinant factor when all other conditions are equal, 
when there are two sets of parameters with about the same Dynamic Range, 
the one which consumed less energy will be favored, and here, 
the measure of energy consumption is the number generated spikes \cite{Attwell2001}.

I stimulated the network with a set of stimuli that were equally spaced in the logarithmic scale, 
separated by $1$ stop ($\times 2$) intervals, similar to FIG. \ref{fig:DynamicRange}.
I ran the simulation for a given time ($1024 ms$) 
and then I calculated Dynamic Range over the range that Just-Noticeable Difference was smaller than 1 stop. 

To be precise, I used a sequence of $n$ stimuli plus the resting state (no stimulus), 
so the sequence includes $n+1$  members $s \in (s_0,  s_1, s_2,  \dots, s_n)$. 
Except $s_0$, the resting state, the difference between $i$-th and $(i+1)$-th stimulus is one stop: $s_{i+1} = 2*s_i$.
Each stimulus $s_i$ evokes a response $r(s_i) \in R$, 
where $R = (r(s_0), r(s_1), r(s_2), \dots, r(s_n))$. 
Now, the Dynamic Range is simply the number of elements in the largest subsequence of $R' \subset R$ that fulfills the expectations, 
\begin{enumerate}
\item For any $r(s_i), r(s_j) \in R'$, we should have $s_i < s_j \Rightarrow r(s_i) < r(s_j)$. 
It means that $R'$ is strictly monotonic and it satisfies the constrain on the Just-Noticeable difference.
\item For any $s_i \in R'$, the network must return to resting state in acceptable time. 
This satisfy the Energy Consumption constrains.
\end{enumerate}
For example, if stimuli $(0, 1, 2, 4, 8, 16)$ leads to the responses $(0, 0, 2, 3, 4, 4)$ 
then the subsequence $(0, 2, 3, 4) \subset R$ is the largest subsequence of $R$ which meets all the conditions; 
Therefore, the dynamic range is $4$. 
In this way the minumum dynamic range can be 1 and its maximum can reach the total number of stimuli. 

\begin{figure}
\centering
\includegraphics[width=8cm]{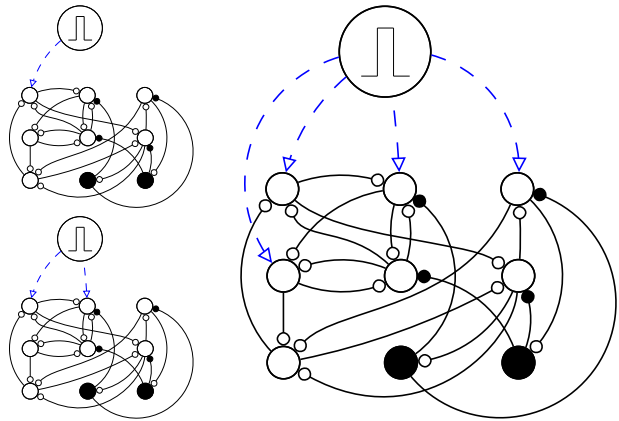}
\caption{A simplified schema of simulation setup. 
Here there are just $N=9$ neurons ($N=1000$ in the actual simulation), 
in which  $N_e = 7$ are excitatory neurons (circles) and $N_i = 2$ are inhibitory neurons (disks). 
The network is stimulated three times, 
each time $s \in (1, 2, 4)$ neurons are stimulated (up left, down left and right respectively), 
with a pulse, in a way that they fire once at time $t=0$ and then they are like any other neuron in the network.}
\label{fig:Network}
\end{figure}

In this study I assumed response $r(s)$ 
as the maximum firing rate of population of excitatory neurons in response to the stimulus $s$. 
But the paradigm of this study is general and can be applied to any neural coding schemes. 

I stimulated some of the neurons with an impulse, I forced $s_i$ excitatory neurons to fire once at time $t=0$, 
then I monitored the propagation of the impulse. 
Here the number of excited neurons represents the magnitude of the stimuli $s_i$. 
This is very similar to Optogenetics stimulation, 
in which, a laser pulse excites a specific population of neurons \cite{Toettcher2011a, Liu2012a, Peron2011, Lim2012}.

I used a {\it simple compartment} {\it Izhikevic model} \cite{Izhikevich2003}.  
It can imitate the behavior of Hodgkin-Huxley model, and its computational costs is comparable to the Integrate-and-fire model \cite{Izhikevich2004a}. 
It is based on a system two-dimensional differential equations:
\begin{eqnarray}
\frac{dv}{dt} &=& 0.04 v^2+ 5 v + 140 - u + I \\
\frac{dw}{dt} &=& a (b v - u) \\ 
\end{eqnarray}
and the after spike equation of:
\[
\text{if } v>30 \text{mV, then}
\begin{cases}
    v \leftarrow c \\
    u \leftarrow u+d
\end{cases}
\]
all units are in $mV$ and $ms$. Here $v$ is the membrane potential, 
$I$ is the input current  and $u$ is a slow variable that imitate the effect of slow moving ions in the cell, like Calcium ions. 
With a proper values of parameters $a, b, c$ and $d$, 
it can model regular spiking neurons, fast spiking neurons and it includes adaptation. 
In our simulaton excitatory neurons are assumed to be regular spiking, inhibitory neurons are assumed to be fast spiking. The values of parameters and the initial values of $u$ and $v$ are set as follow: \\

\begin{tabular}{|l|c|c|c|c|c|c|}
\hline
&a&b&c&d&u&v\\ \hline
Regular Spiking & 0.02 & 0.2 & -65.0 & 8.0 & -14 & -70 \\ 
Fast Spiking &‌ 0.10 & 0.2 & -65.0 & 2.0 & -14 & -70 \\ \hline
\end{tabular}\\

My network has a random graph topology \cite{Erdos1960}. 
In a random graph, every two neurons are connected by probability $p$, called sparseness. 
This is one of the simplest topology for a network or graph.

I simulated networks of $N=1000$ neurons with \[N_I \in (100, 200, 300, 400, 500)\] inhibitory neurons. 
The neurons were connected by a random graph topology, 
with the sparseness   \[p \in (0.01, 0.02, 0.04, 0.08, 0.16).\]

On the other hands, the synaptic weights are chosen randomly, 
but in a way to cover the region of interest in the $w_e$, $w_i$ plane. 
In this way, it is always possible to add new points later to improve results. 
The new points could be just another pair of connection weights$(w_e, w_i)$ on the same random netwrok,
or they could be synaptic weights of a whole new realization. 

I stimulated each network 10 times, \[s \in (0, 1, 2, 4, 8, 16, 32, 64, 128, 256 )\] 
so in each trial, $s$ neurons  fire at $t=0$. 
Then I watched the network for $1024 ms$.
At the end, I calculated Dynamic Range of the network. 
I repeated the whole process for few realization of random networks and and then the median of all data in each hexagonal is calculated and displayed (FIG. \ref{fig:Network}).

The whole process needs a great amount of computational power. 
That is the reason I used {\it NeMo} (http://nemosim.sourceforge.net/), 
a neural simulator software that runs on GPU \cite{Fidjeland2009, Fidjeland2010}, 
but for the pilot study, I used {\it Brian} (http://briansimulator.org/) \cite{Goodman2008}. 
Both of them are open source and available on their websites. 

\section{Results}
\begin{figure}
\centering
\includegraphics[width=5cm]{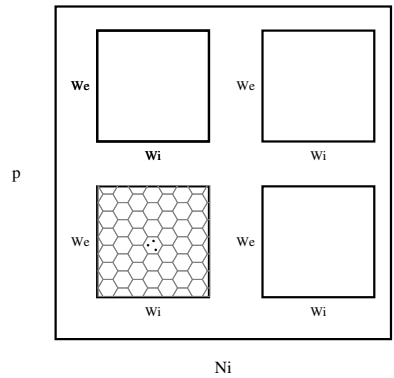}
\caption{The layout of phase diagram, (FIG.\ref{fig:erate} and others), 
Here, $p$ is the sparseness of connections, 
$N_i$ is the number of inhibitory neurons, 
$w_e$ is the weight of excitatory synapses (a positive number) and $w_i$ is the weight of inhibitory synapses (a negative number).
The plane of $w_e$ and $w_i$ is divided into hexagonal bins. 
The color of each hexagon is defined by the aggregated value of data points inside that hexagon.
For aggregation, unless specified, median value of points are used.
If a hexagon is empty, then its color is white.
}
\label{fig:template}
\end{figure}
\begin{figure}
\includegraphics[width=16cm]{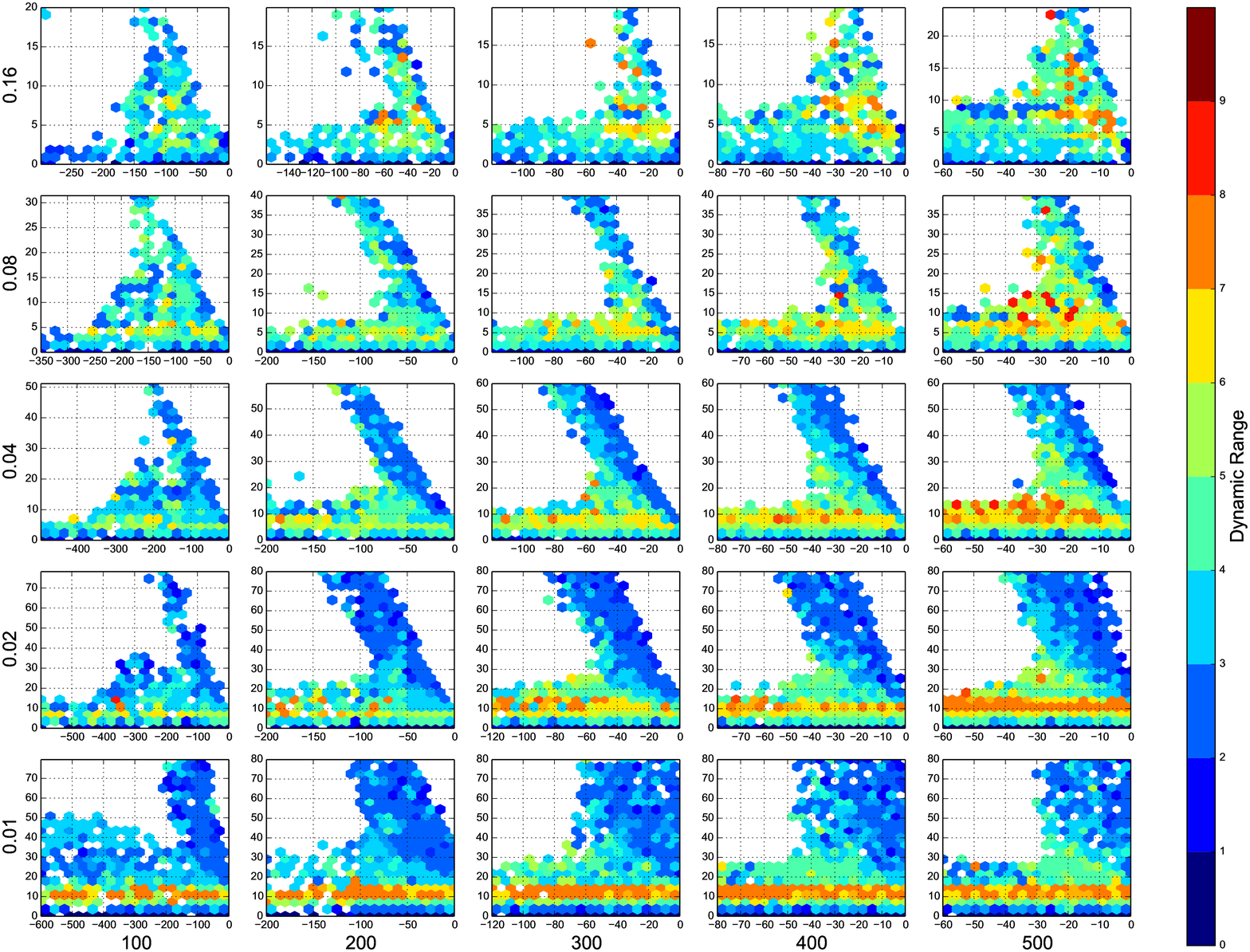}
\caption{The phase diagram of dynamic ranges, 
dark red means high dynamic range (good), white means there was not any simulation data at that bin, 
all the data were unreliable, or the the network did not come back to its resting state before 100 ms of simulation.    
The information are arranged according to the template of FIG.\ref{fig:template}.}
\label{fig:erate}
\end{figure}
\begin{figure}
\includegraphics[width=16cm]{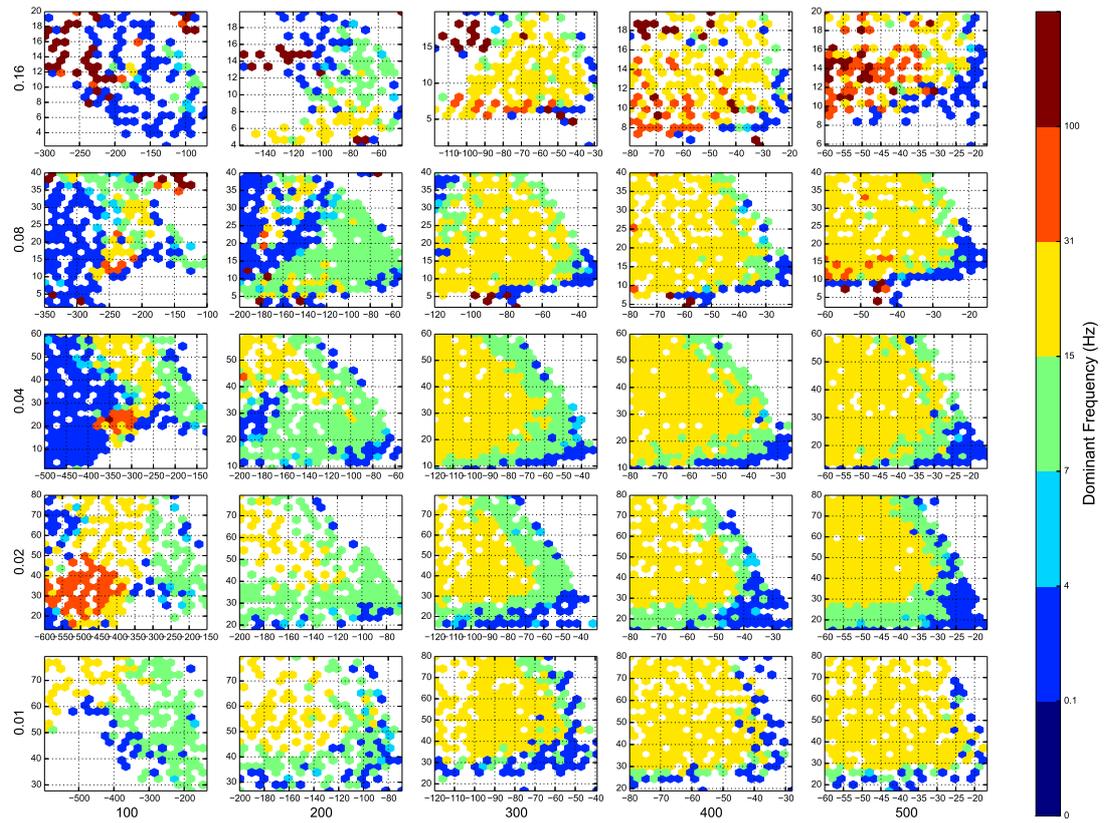}
\caption{The phase diagram of oscillations,
the dominant frequency of population rate of the network according to FFT. 
Colors are associated to brain waves: 
$\delta (0.1-4 Hz), \theta (4-7 Hz), \alpha (7-15 Hz), \beta (15-31 Hz)$ and $\gamma(31-100 Hz )$,
and color of a bin shows the median of all trials in that bin.
Here white means there was not any simulation data at that bin, 
all the data were unreliable (they fired at the maximum possible population firing rate), 
or the network did not show any activity after 100 ms of simulation.
The information are arranged according to the template of FIG.\ref{fig:template}.}
\label{fig:oscillations}
\end{figure}
\begin{figure}
\includegraphics[width=16cm]{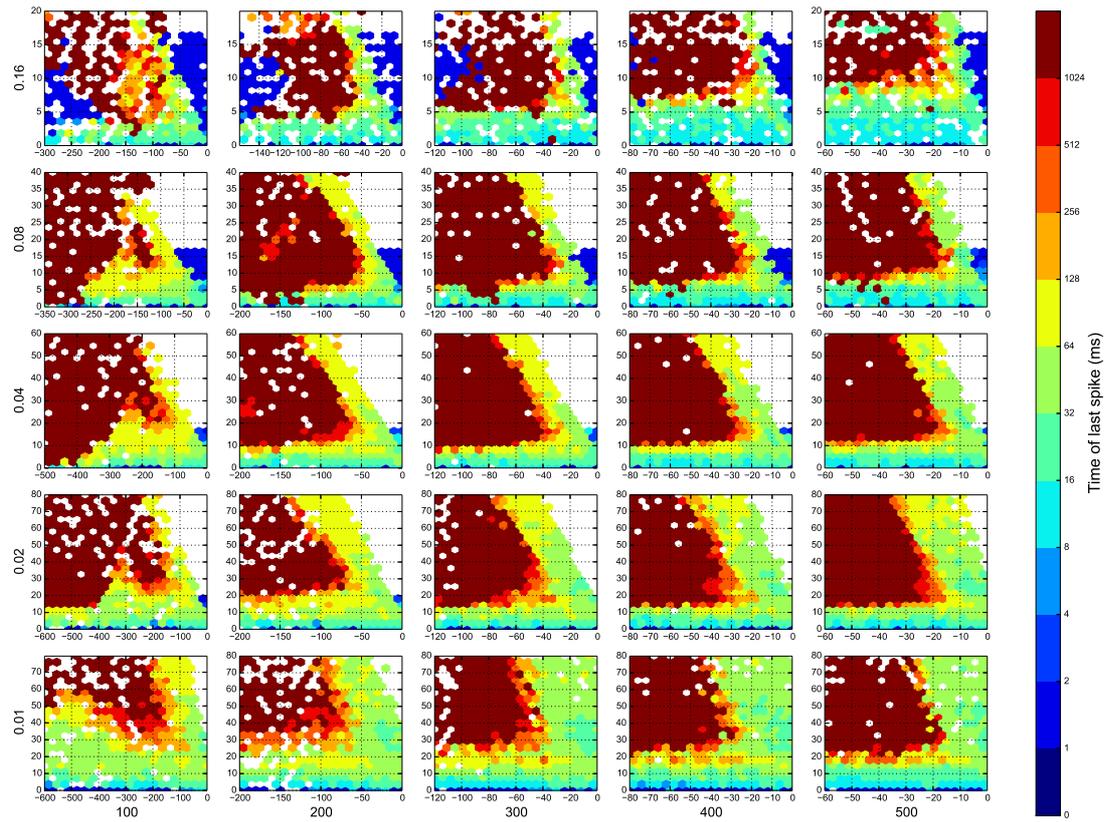}
\caption{The time of last spike in each trial.  
    Dark red means the simulation ended before the network returns to its resting state.
    Here white means there was not any simulation data at that bin, 
    or all the data were unreliable.
    The information are arranged according to the template of FIG.\ref{fig:template}.}
\label{fig:last-spike}
\end{figure}
\begin{figure}
\includegraphics[width=16cm]{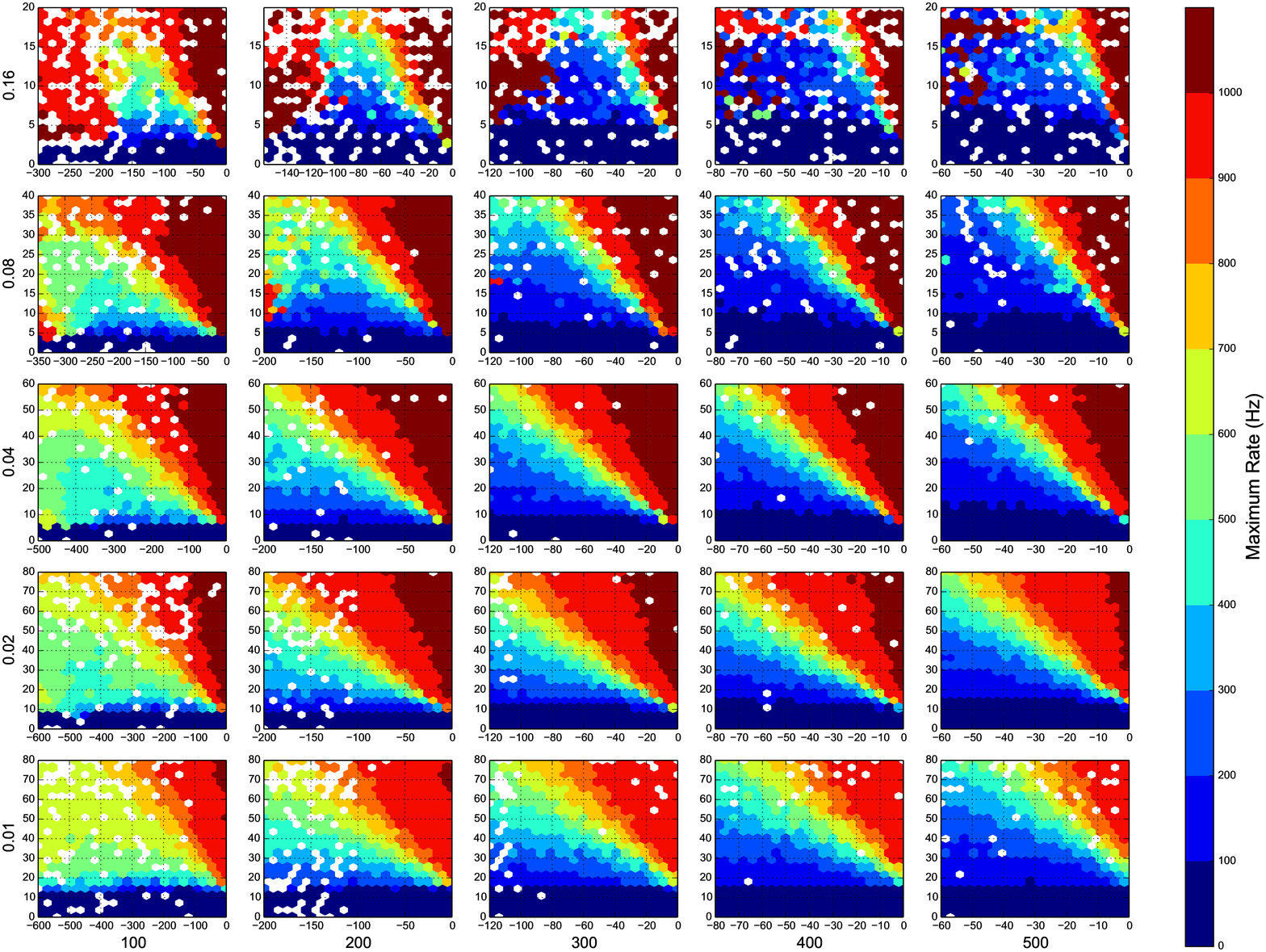}
\caption{Maximum firing rate in each trial.
Here white means there was not any simulation data at that bin.
The information are arranged according to the template of FIG.\ref{fig:template}.}
\label{fig:max-rate}
\end{figure}
\begin{figure}
\includegraphics[width=16cm]{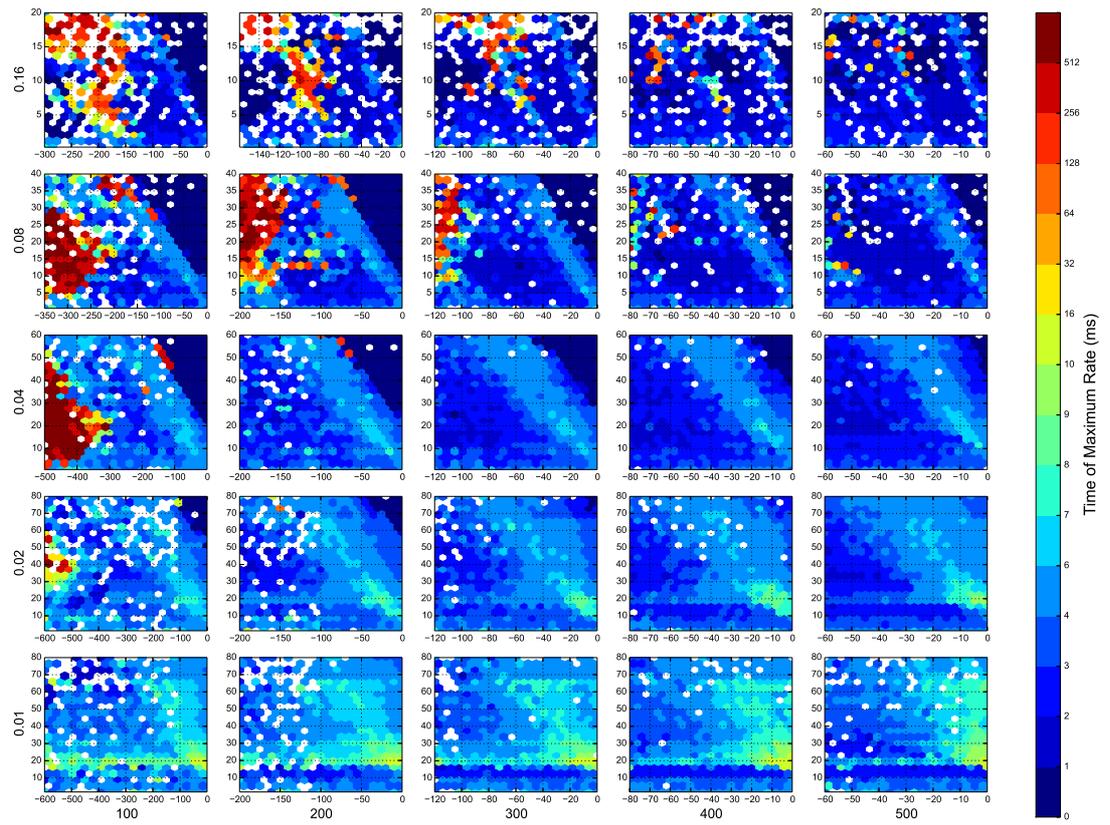}
\caption{The time that maximum firing rate happens in each trial. 
Here white means there was not any simulation data at that bin.
The information are arranged according to the template of FIG.\ref{fig:template}.}
\label{fig:time-of-max-rate}
\end{figure}

I wanted to present the  results as a function of 4 other parameters, 
for that I needed to map a 4-D space to a 2-D space. 
Here I used the template in figure \ref{fig:template}, 
to present the dynamic ranges in figure \ref{fig:erate}, 
oscillations in figure \ref{fig:oscillations} , 
duration of activities in figure \ref{fig:last-spike}, 
maximum achieved firing rate in each trial in figure \ref{fig:max-rate}
and the time of achieved maximum rate in figure \ref{fig:time-of-max-rate}
First I made the plots of dynamic range in the space of synaptic weights -- $w_e$ and $w_i$. 
Then I put these plots next to each other according to sparseness of connections $p$ and number of inhibitory neurons $N_i$.


\begin{figure}
\centering
\includegraphics[width=4cm]{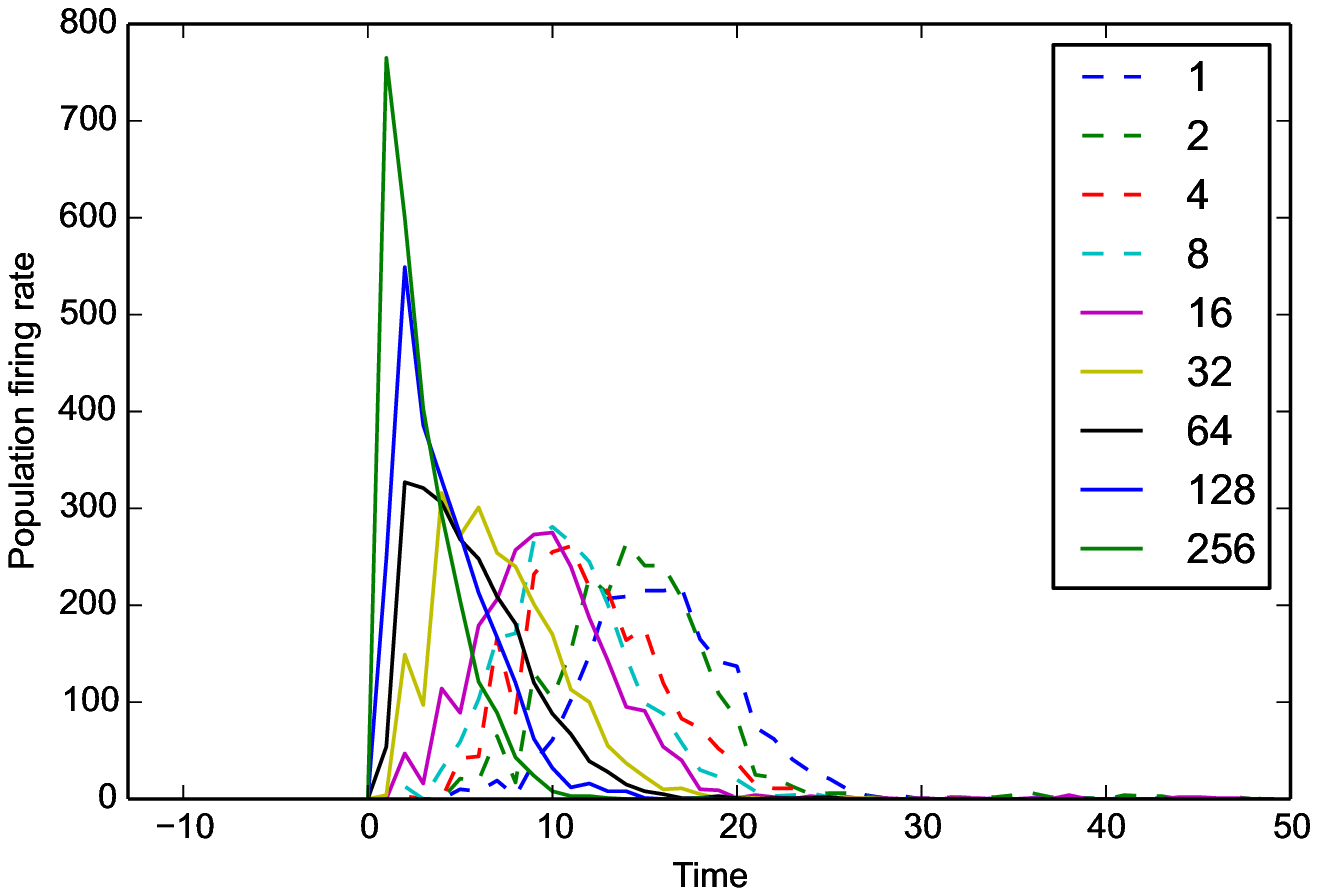}
\includegraphics[width=4cm]{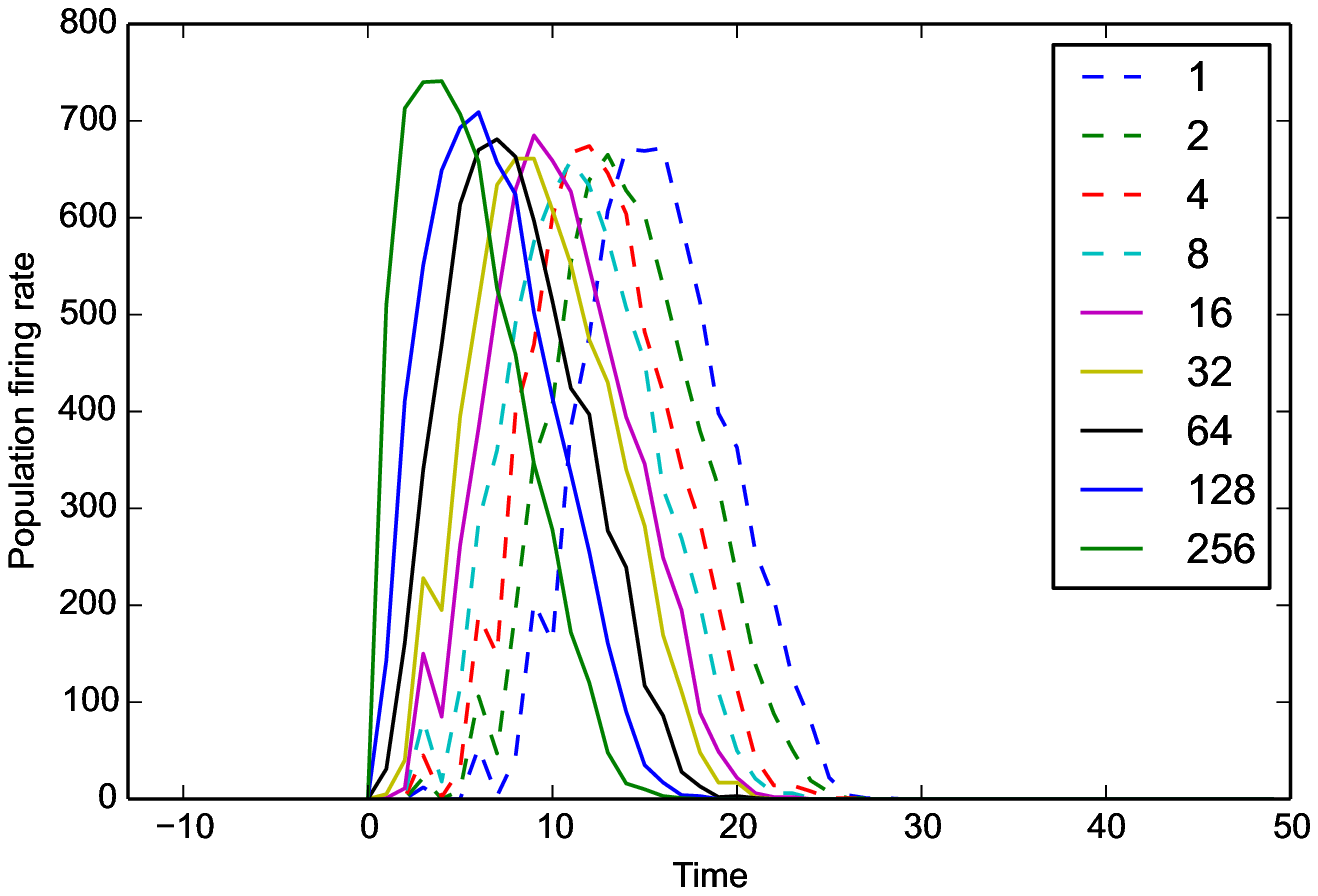}
\includegraphics[width=4cm]{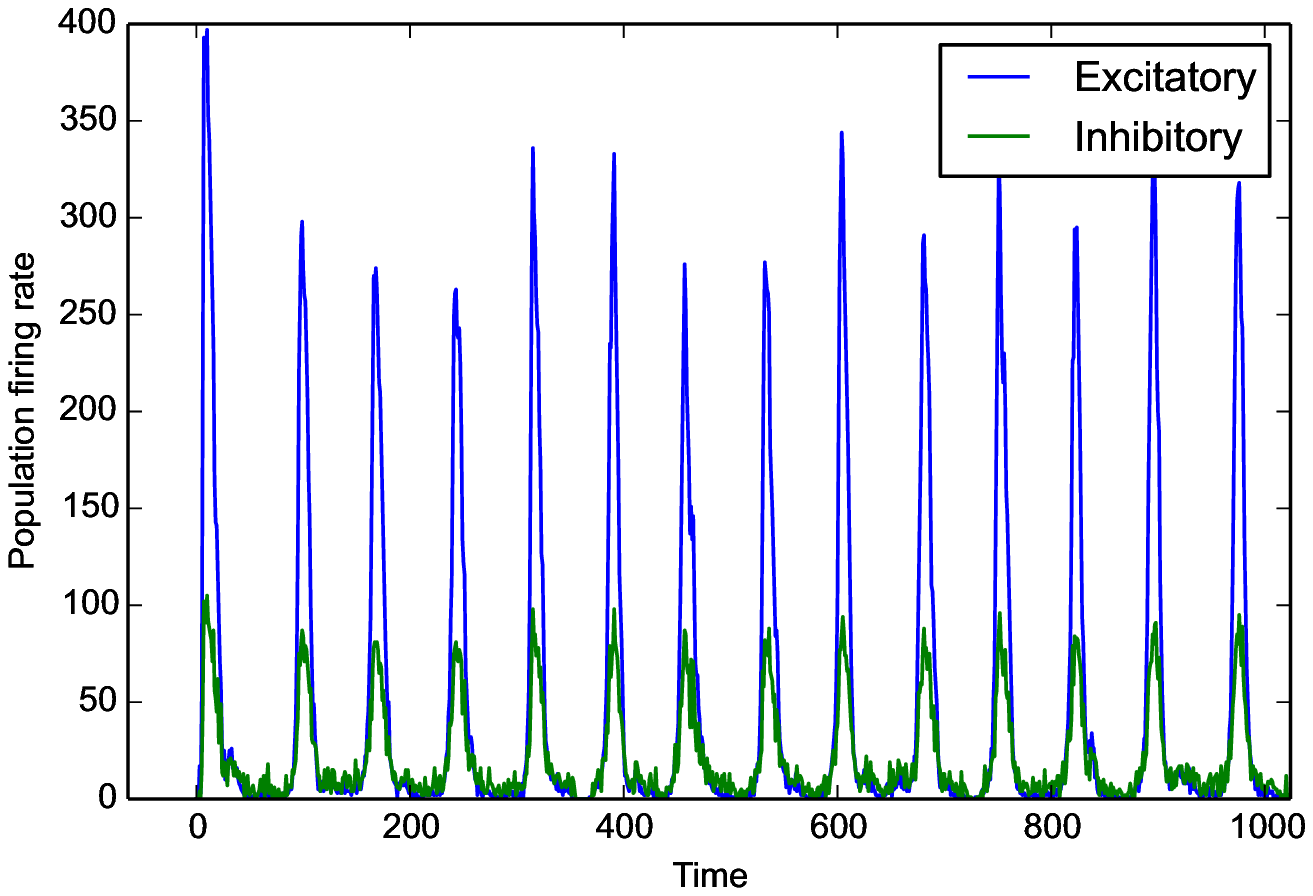}\\
\includegraphics[width=4cm]{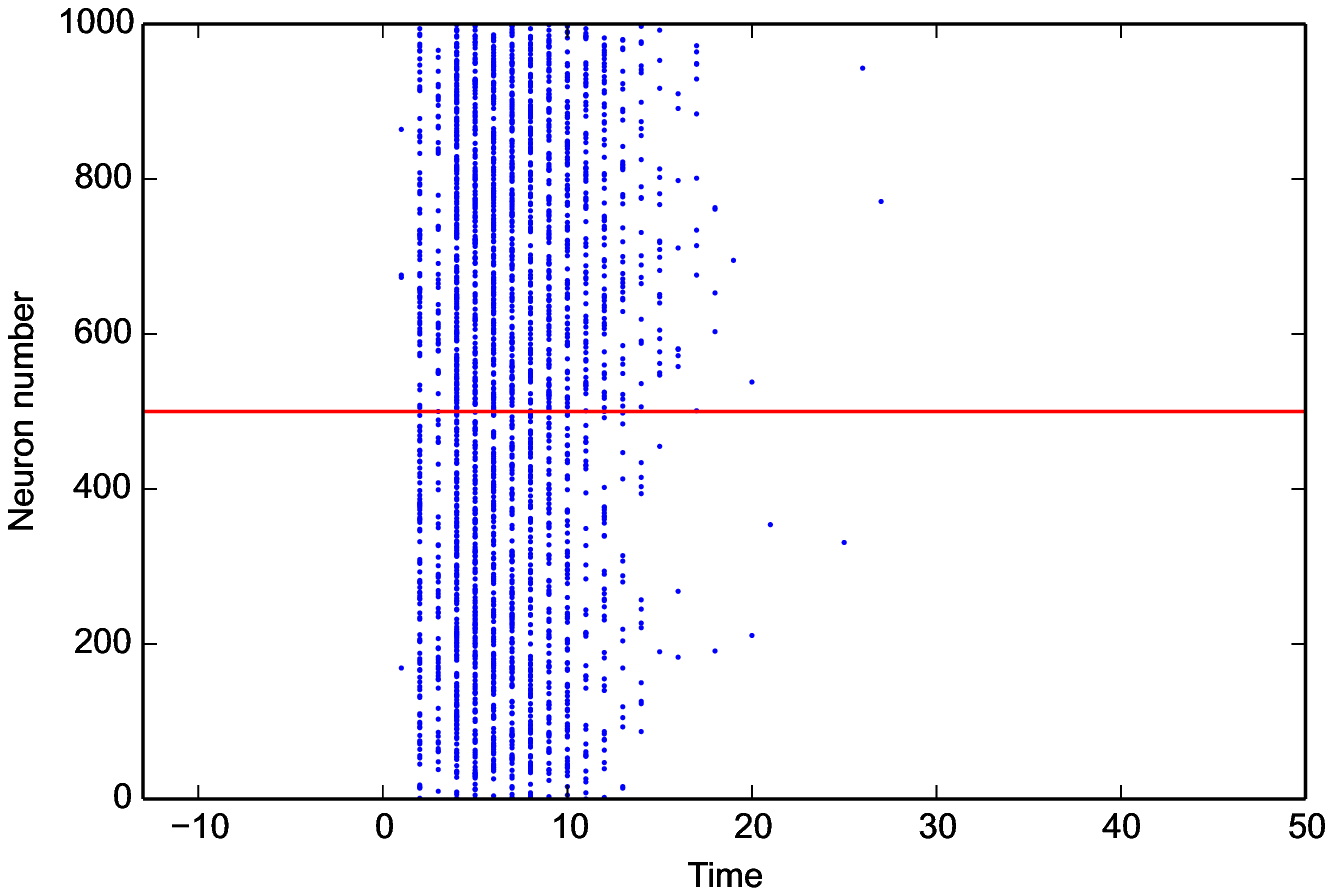}
\includegraphics[width=4cm]{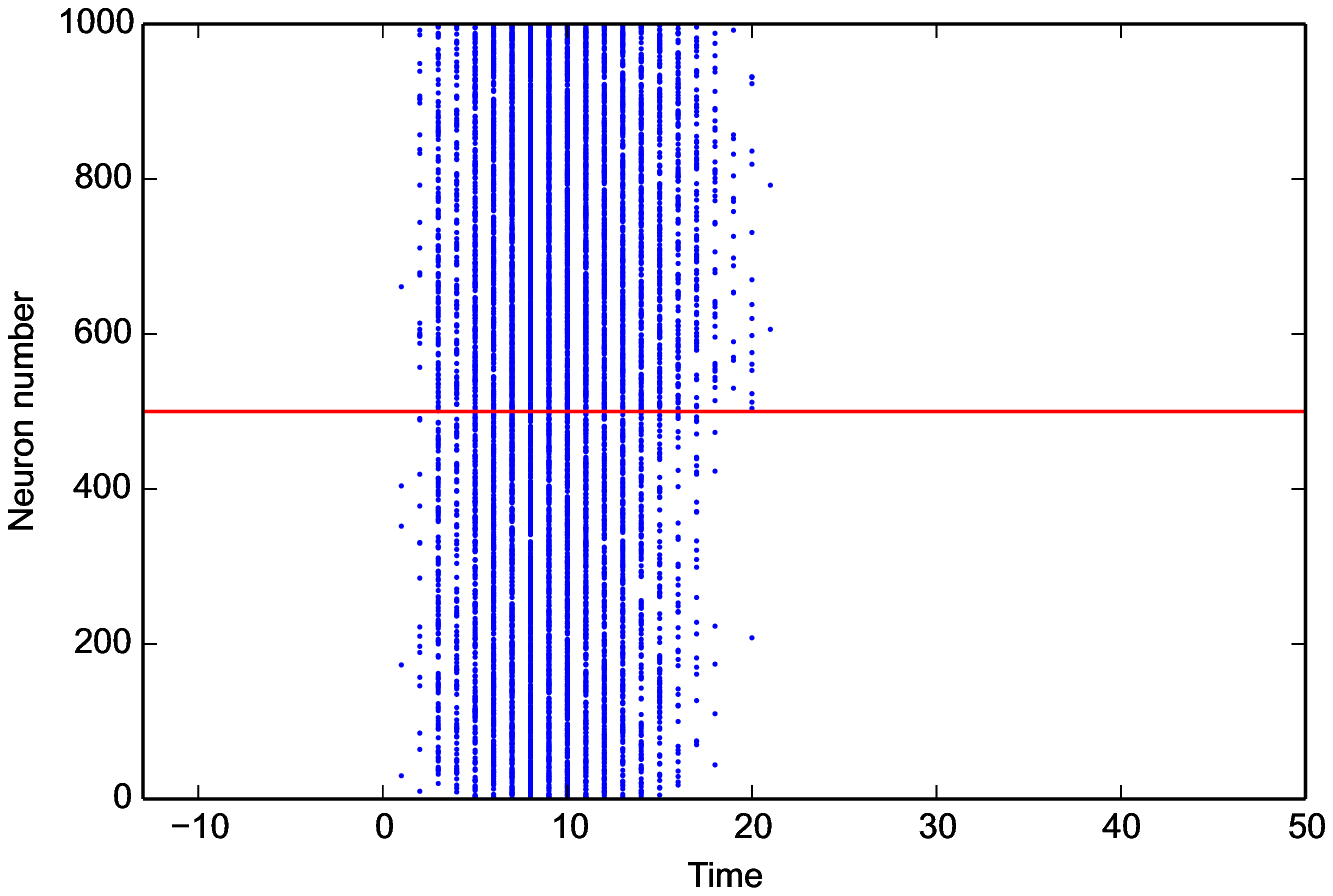}
\includegraphics[width=4cm]{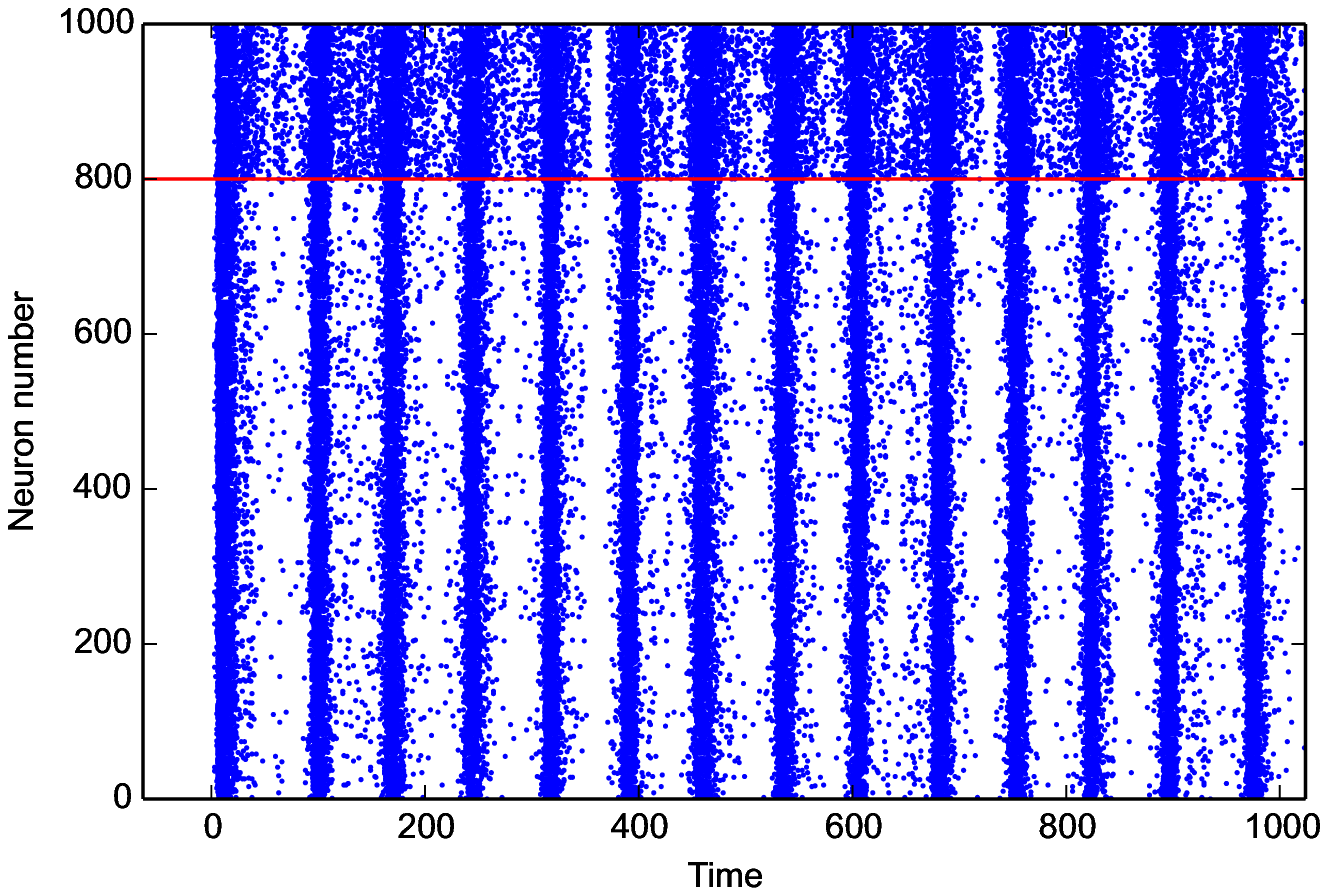}
\caption{ An example of fast respoding network with a good dynamic range (left),
    poor dynamic range (middle) and 
    an example of oscillatory network (right).    
    Population rates as a function of time (top).
    The spiking activity of neurons (bottom). 
    The red line separates excitatory neurons (below the line) from inhibitory neurons (above the line). 
}
\label{fig:examples}
\end{figure}
\section{Discussion}
In figure \ref{fig:erate}, top right corner of each figure is empty (white). 
This is where the excitatory neurons dominates and the network riches its maximum firing rate, 
this region is visible in figure \ref{fig:max-rate} as the darkest shade of red.
The left side of each graph, where the inhibitory sub-network dominates, 
is also empty, but this time because networks did not return to resting states.
More detail about the length of activity demonstrated in figure \ref{fig:last-spike}. 

A narrow band separate this two regions, on this band, the inhibitory and excitatory network are balancing out each other. 
As we increase the number of connections, this band gets narrower, the delicate balance can be easily disturbed.
The dynamic ranges on this band is low or moderate.  

At the bottom of each graph, there is another region with valid data points. In this region, 
excitatory sub networks are not strong enough to saturate the system, 
so we have a good dynamic range regardless of the inhibitory synaptic weights.
Nevertheless, stronger inhibitory sub networks improves the dynamic range.   

For the calculation of dynamic range, 
only trials who returned to resting state have been used. 
But in many other trials the network sustained its activity up to the very last millisecond of simulation (FIG \ref{fig:last-spike}). 
These networks shows interesting oscillatory behaviors, 
that can be seen in figure \ref{fig:oscillations} and figure \ref{fig:examples}.
The oscillations are mostly $\alpha$ and $\beta$ waves, 
with fewer case of $\gamma$ waves. 
There are also $\delta$ and $\theta$ waves, 
but the simulation time of 1 second may not be enough to study these waves. 
We also didn't consider the effect of $GABA_B$ receptors, 
which may be dominant factor in slow rhythms oscillations like $\delta$ and $\theta$ waves.

The oscillation and synchrony of spiking neural networks are studied intensivly. 
Few examples are \cite{Maheswaranathan2012, Augustin2013, Sadeghi2014, Akam2010}. 
In most of these works, the neurons are intrinsically oscillatory or 
there is an external force -- a current, some noise or a few poison spike generators. 
But here, the networks are stimulated with only a pulse, and then they are left alone, 
even though there is not any spontaneous activity implemented in the model, yet, 
the networks show sustained oscillations. 

The oscillation are mostly $\beta$-waves, and they happen in the region of dominating inhibitory sub networks.
As we get to the balanced inhibitory-excitatory band, the oscillations become $\alpha$-waves.

It seems that the results are not very sensitive to the parameters, 
We can have good dynamic range over some value of parameters, 
and we can have neural oscillators in different settings. 
These finding confirms former studies \cite{Prinz2004, Marder2007, Marder2011, Marder2011a, Gutierrez2013}.

The results of this paper, specially figure \ref{fig:erate} and figure \ref{fig:oscillations}, 
serves as a starting point to decide about parameters in a neural simulation, 
and they may help to find the networks with desired behavior.  



This work by no mean is complete, I only studied the effect of four parameters, 
in just one model, over dynamic range  and oscillations. 
Many other parameters could be included, 
like network topology, synaptic delay and distribution of weights or connections. 
Also it is interesting to know if my results holds for simpler models, like Integrate-and-fire. 

At the end I suggest that any numerical simulation of neural network should be 
accompanied with such a phase diagram, it demonstrate the robustness of the results in respect to the parameters. 
One good example is the the work of \cite{Roy2013}, 
where they made such diagram to compare experimental data of 
orientation selectivity index of mouse primary virtual cortex to their model. 



\section{Acknowledgment}
The research presented in this paper was carried out on the High Performance Computing Cluster supported by the Computer Science department of Institute for Research in Fundamental Sciences (IPM).

\bibliographystyle{plain}	
\bibliography{SpikingNeuralNetworks}

\end{document}